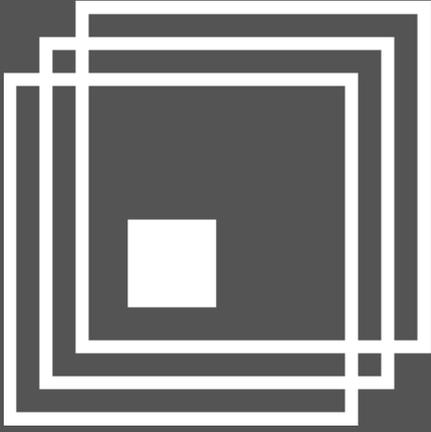

# IOT
# ROADMAP

SUPPORT FOR INTERNET OF THINGS
SOFTWARE SYSTEMS ENGINEERING

**AUTHORS**
**Rebeca Motta** (COPPE/UFRJ)
**Káthia Oliveira** (LAMIH/UPHF)
**Guilherme Travassos** (COPPE/UFRJ)

**PROPOSAL**
What to Consider When Engineering IoT

**FOCUS**
IoT Development Teams

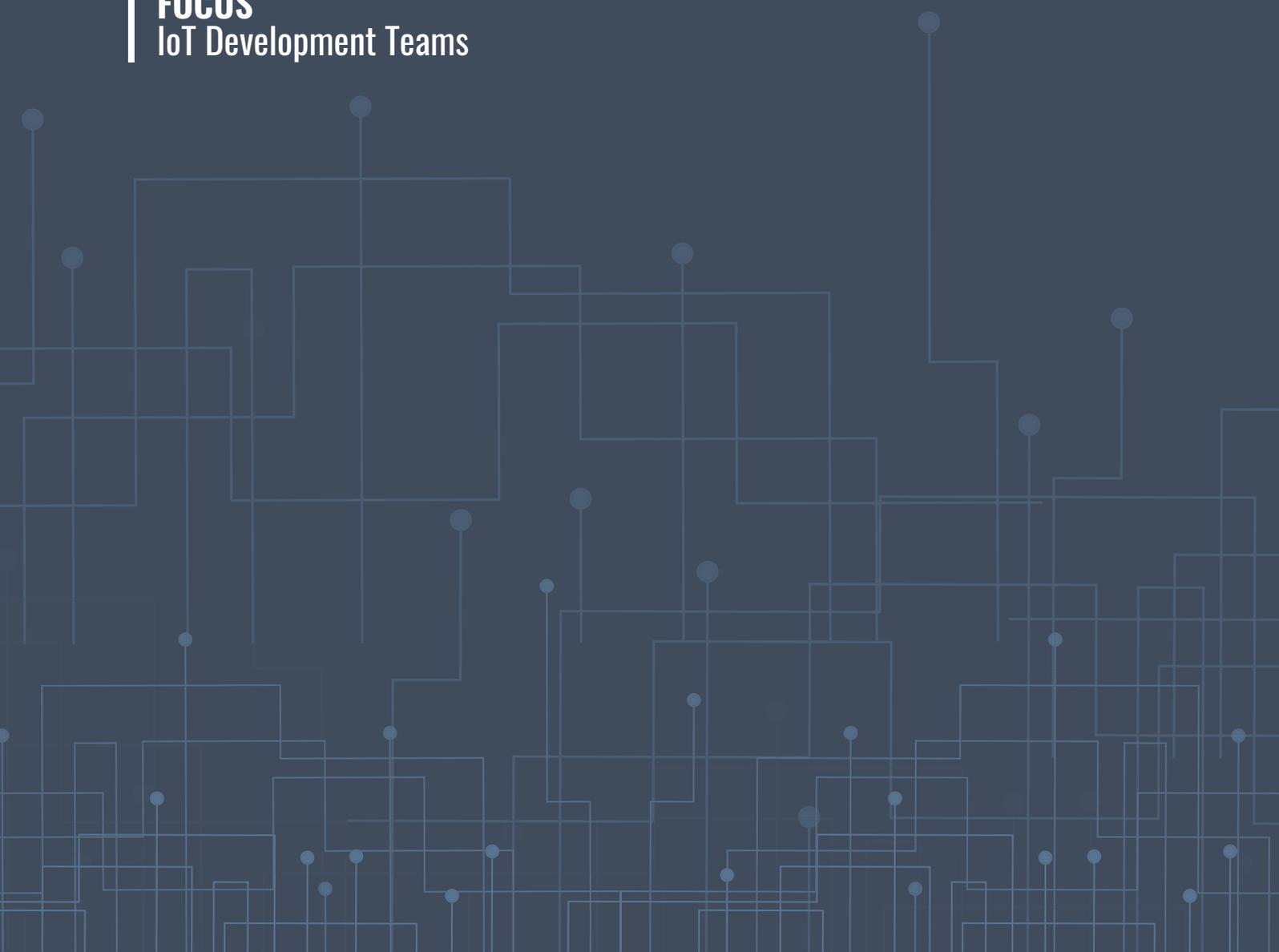

To cite this work please refer to: MOTTA, Rebeca; OLIVEIRA, Káthia; TRAVASSOS, Guilherme. IoT Roadmap: Support for Internet of Things Software Systems Engineering. arXiv preprint arXiv:2103.04969, 2021.

# ABOUT THE IoT ROADMAP:

## THE RESEARCH

The present Roadmap is performed in the context of a Ph.D. research in collaboration between the Experimental Software Engineering Group, from the Systems Engineering and Computer Science Program of the Federal University of Rio de Janeiro and the Laboratory of Industrial and Human Automation Control, Mechanical engineering and Computer Science (LAMIH UMR CNRS 8201) in the Universite Polytechnique Hauts-de-France.

## THE PROPOSAL

The Roadmap resulted from an investigation on the particularities of new IoT applications. We consider current challenges in their development and multidisciplinarity. The Roadmap organizes the concepts and evidence gathered from different experimental studies. It comes to support IoT software systems' definition, with specific items for the project team to discuss and define the essential aspects of specifying, designing, and implementing an IoT application.

# HOW TO USE IT

The team should **(1) read** the Roadmap to encourage discussions of the details related to each facet. They should **(2) consider** each question before following the recommendations. This way, the understanding of the items is aligned among all the team. The Roadmap does not aim to replace everyday activities in the development or the original methods in more traditional software projects. Nevertheless, the Roadmap can be **(3) combined** with the existing methods and technologies already in use. Last, it is to **(4) performe** their established strategy for the project.

The Roadmap can be used iteratively, following the project cycles. With this, it is possible to minimize the project uncertainty. All stakeholders can use it as a guide to support discussions and decision-making for an action plan for the development.

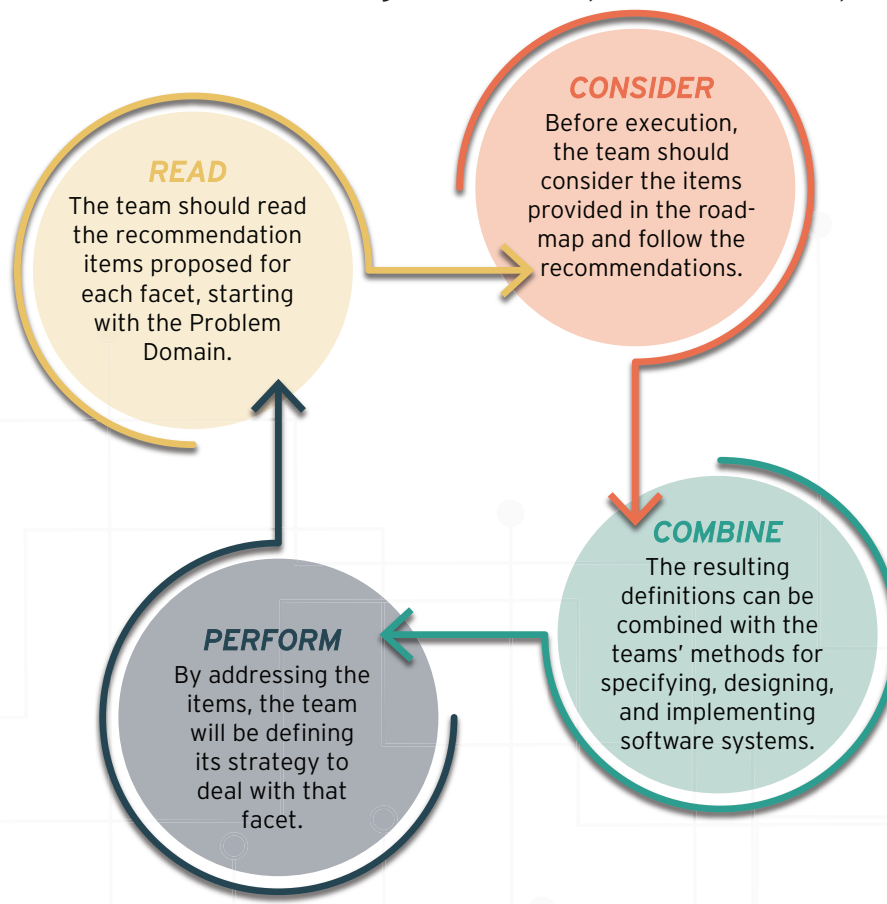

The items are grouped in categories. Each item can be marked as **Done** - if it is already completed, **To Do** - if it is an activity for the next phases, and **Not Applicable (N/A)** - if it is not in the project plan.

**Please pay attention** to *Cross-cutting items* since they can evolve and change throughout the project. These items are marked with: 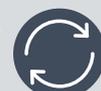

# IoT ROADMAP CONCEPT: THE PHASES

## *Based on the Body of Knowledge to Advance Systems Engineering*

Systems engineering presents "an interdisciplinary approach and means to enable the realization of successful systems. Successful systems must satisfy the needs of their customers, users, and other stakeholders". It concerns activities to discover, create and describe a system to satisfy an identified need. The activities are grouped and described as general processes that cover build artifacts, decisions for concept definition, stakeholders' needs and requirements, and preliminary operational concepts. It considers a generic life cycle with the following phases:

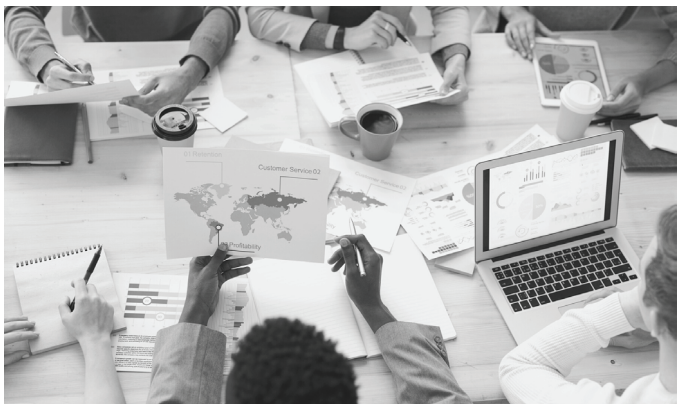

### CONCEPT DEFINITION

Where there is a decision to invest resources in a new or improved engineered system. Consists of developing the concept of operations and business, determining the key stakeholders, requirements and the system desired capabilities.

### SYSTEM DEFINITION

Evolution and formalization of the requirements to be sufficiently well defined. It provides a basis of a system realization considering the architecture and design, compatibility, and feasibility of the resulting system definition.

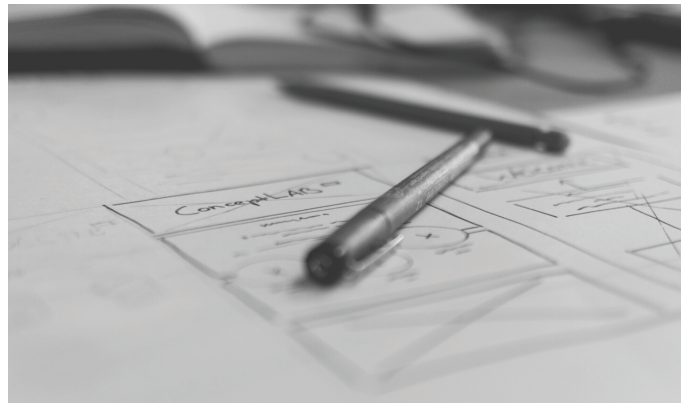

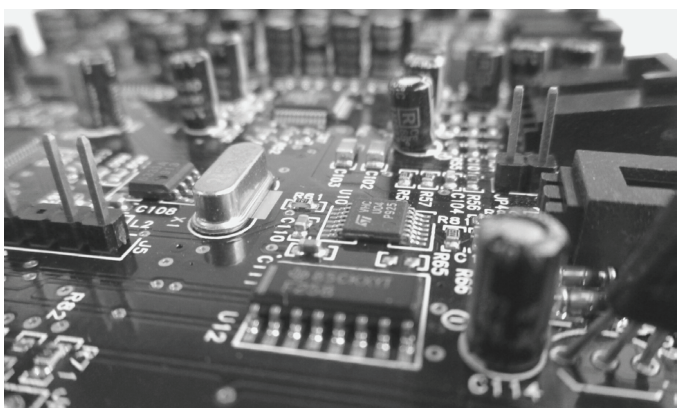

### SYSTEM REALIZATION

Aimed to deliver operational capability with the construction as well as the integration of the elements. Verification and validation, and preparing for the production, support and utilization activities are included.

# IoT ROADMAP CONCEPT: THE FACETS

The Roadmap aims to address some of the IoT's existing challenges and its multidisciplinarity. The facets represent different disciplines and knowledge areas involved in IoT. Meaning "one side of something many-sided" (Oxford Dictionary), "one part of a subject, a situation that has many parts" (Cambridge Dictionary), representing the multidisciplinarity required in such systems.

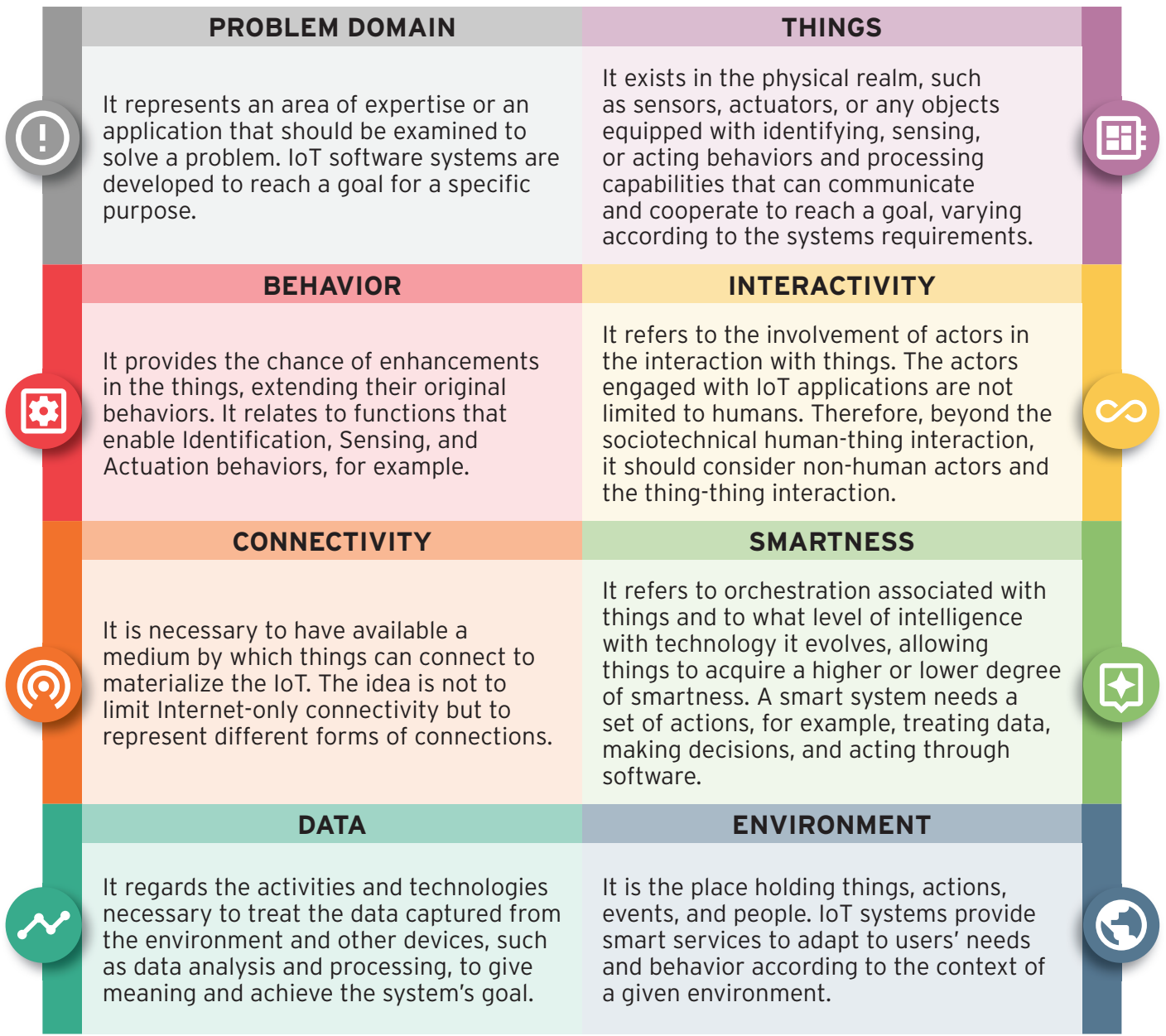

**PROBLEM DOMAIN**

It represents an area of expertise or an application that should be examined to solve a problem. IoT software systems are developed to reach a goal for a specific purpose.

**THINGS**

It exists in the physical realm, such as sensors, actuators, or any objects equipped with identifying, sensing, or acting behaviors and processing capabilities that can communicate and cooperate to reach a goal, varying according to the systems requirements.

**BEHAVIOR**

It provides the chance of enhancements in the things, extending their original behaviors. It relates to functions that enable Identification, Sensing, and Actuation behaviors, for example.

**INTERACTIVITY**

It refers to the involvement of actors in the interaction with things. The actors engaged with IoT applications are not limited to humans. Therefore, beyond the sociotechnical human-thing interaction, it should consider non-human actors and the thing-thing interaction.

**CONNECTIVITY**

It is necessary to have available a medium by which things can connect to materialize the IoT. The idea is not to limit Internet-only connectivity but to represent different forms of connections.

**SMARTNESS**

It refers to orchestration associated with things and to what level of intelligence with technology it evolves, allowing things to acquire a higher or lower degree of smartness. A smart system needs a set of actions, for example, treating data, making decisions, and acting through software.

**DATA**

It regards the activities and technologies necessary to treat the data captured from the environment and other devices, such as data analysis and processing, to give meaning and achieve the system's goal.

**ENVIRONMENT**

It is the place holding things, actions, events, and people. IoT systems provide smart services to adapt to users' needs and behavior according to the context of a given environment.

The problem domain will direct and contextualize how the other facets will be derived, implemented, and managed. The roamap helps to go from the problem domain to a software solution, considering all the facets as part of the same solution, one related to the other aiming at the completion.

# IoT ROADMAP CONCEPT: THE ELEMENTS

*The structure of the Roadmap combines phases, facets, and items.*

The **phases** organize the engineering life cycle through time. It goes from the need for an IoT product to the construction of the product itself.

In IoT, several areas are intertwined to achieve a solution. Therefore the phases should be considered in a multi-faceted way to address the IoT requirements in a multidisciplinary fashion with the **facets**.

Each facet comprises a set of **items**, which represent activities, definitions, and recommendations for the project team to achieve the desired solution.

| PHASES | FACETS | ITEMS |
|---|---|---|
| *A generic life cycle with three phases* | *Disciplines and knowledge areas involved in IoT* | *Recommendations for IoT particularities* |
| From conception to realization, the phases organizes the activities to transform requirements into a deliverable IoT software product. | IoT software-based solutions involve seven facets that require different skills and technologies to go from the problem to the solution domain. | Activities to support the project team to discuss and define aspects related to the specifying, designing, and implementing an IoT application. |

# IoT ROADMAP CONCEPT: THE ELEMENTS

*The structure of the Roadmap combines phases, facets, and items.*

Early-stage decisions guide the product being specified, designed, and engineered. They clarify the overall scope and establish a basic understanding of the problem, the type of solution desired, and the team that will oversee the solution.

The Roadmap contemplate the three phases with specific recommendation items for each facet. The items can be assigned in different status for to do, done or not applicable according to the project definition and evolution.

\* Some **crosscutting items** 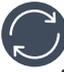 can be revisited and evolved throughout the project, considering new information that can update the implementation. The revision and changes of the phases should follow the teams' internal direction for development cycles, such as sprints.

\* It is possible to add comments and details in the **Reference Field** to  better organize project information and help with the traceability.

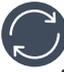

| | PROBLEM DOMAIN | | | THINGS | | | BEHAVIOR | | | INTERACTIVITY | | | CONNECTIVITY | | | SMARTNESS | | | DATA | | | ENVIRONMENT | | |
|---|---|---|---|---|---|---|---|---|---|---|---|---|---|---|---|---|---|---|---|---|---|---|---|---|
| | to do | done | n/a | to do | done | n/a | to do | done | n/a | to do | done | n/a | to do | done | n/a | to do | done | n/a | to do | done | n/a | to do | done | n/a |
| Concept Definition | 10 | 5 | 3 | 12 | 2 | 5 | 7 | 1 | 9 | 3 | 0 | 0 | 4 | 1 | 0 | 2 | 0 | 3 | 8 | 0 | 1 | 4 | 0 | 3 |
| System Definition | 13 | 2 | 3 | 4 | 10 | 5 | 5 | 3 | 9 | 2 | 1 | 0 | 2 | 2 | 0 | 0 | 2 | 3 | 4 | 4 | 1 | 1 | 3 | 3 |
| System Realization | 0 | 15 | 3 | 0 | 14 | 5 | 0 | 8 | 9 | 0 | 3 | 0 | 0 | 4 | 0 | 0 | 2 | 3 | 0 | 8 | 1 | 0 | 4 | 3 |

The IoT Roadmap starts in the next section and by following its recommendation it is possible to minimize the uncertainty of IoT project development.

# IoT ROADMAP PRACTICAL
*To be used to support problem understanding and the formal definition for conception and realization towards an IoT solution.*

## PROBLEM DOMAIN

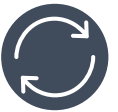

**1. Define the objective and motivation for the IoT project.** An IoT-based solution is provided for a particular goal based on a real problem and motivation. From the data we observed, the inspiration behind the solution could affect how the problem is addressed.

| | TO DO | DONE | N/A |
|---|---|---|---|
| Define the problem domain, highlighting the need for IoT solution (such environmental control with real-time actuation). | ◉ | ◉ | ◉ |
| Define the system goal, highlighting the IoT characteristics (such as communication in real-time, wider range and scale, and remote control). | ◉ | ◉ | ◉ |
| Establish the problem motivation for using IoT technology (such as optimization of resources and requirement for less human intervention). | ◉ | ◉ | ◉ |

**2. Define IoT system behavior.** Define the basis for the project, defining what the stakeholders – users, things, developers, actors – need from it and what the system must do to satisfy this need. Be well understood and defined by everybody, capture the idea of the product.

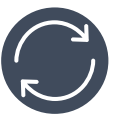

| | TO DO | DONE | N/A |
|---|---|---|---|
| Define high-level IoT requirements (such as the use of sensors and tags to address sensing). | ◉ | ◉ | ◉ |
| Describe high-level IoT behaviors (such as sensing the context of a given environment and actuation in a production line). | ◉ | ◉ | ◉ |
| Identify high-level users, roles, actors (such as external service to contribute with information and users with permissions to adjust the actuation rules). | ◉ | ◉ | ◉ |
| Establish the high-level context of use (such as a during cropping season in a farm, healthy control in a water tank, and a maintenance in a production site). | ◉ | ◉ | ◉ |

**3. Define IoT system limitations.** A specification or technical limitation to achieve some functionality. It refers to what was defined and what the system doesn't do. This limitation can lead to recommendations for improvements.

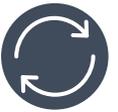

| | TO DO | DONE | N/A |
|---|---|---|---|
| Establish the IoT technical limitations (such as the size of the solution not to disturb the animals on a farm, and the need to be waterproof in a water tank). | ◉ | ◉ | ◉ |
| Establish the IoT functional limitations (such as the system will only capture soil information, not weather information, and act based on the human decision, not automatically). | ◉ | ◉ | ◉ |

**4. Verify existing IoT solutions.** To decide on building or adapting a component, supporting the decision to develop a new system, or being aware of current technologies and available options. Prior research is required to verify existing solutions.

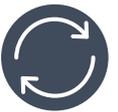

| | TO DO | DONE | N/A |
|---|---|---|---|
| Describe existent IoT systems or products (such as similar products to what is going to be developed). | ◉ | ◉ | ◉ |
| Describe existent technologies for IoT (such as check if any add-on or component is already available). | ◉ | ◉ | ◉ |

**5. Define solution benefits and risks.** The proposed IoT solution can achieve the expected goal and deliver some advantages from other alternatives. However, it can also have a downside and possibility of damage, loss, difficulty, or threats generated from the IoT solution.

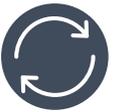

| | TO DO | DONE | N/A |
|---|---|---|---|
| Define the benefits of using the IoT solution (such as immediate assistance due to real-time controlling and less human intervention due to smartness). | ◉ | ◉ | ◉ |
| Describe and implement mechanisms to mitigate the risks (such as to define regulatory compliance and control access both physical and virtual to the IoT solution). | ◉ | ◉ | ◉ |
| Establish the possible risks (such as user and product safety). | ◉ | ◉ | ◉ |

**6. Define strategy for relevant quality characteristics and attributes.** The project should have a clear definition of their quality characteristics (assigned property) and attributes (inherent property) in a compliant way with the specification and general expectations. Establish practices to ensure the overall quality and constraints of the system. From the high-level attributes identified as system goals, refine to manageable and measurable items.

**Some quality characteristics examples retrieved from IoT projects are:** Acceptance, Accessibility, Adaptability, Attractiveness, Automation, Availability, Compatibility, Controllability, Frequency, Integrity, Interoperability, Intrusiveness, Learnability, Mobility, Performance, Precision, Privacy, Range, Reliability, Safety, Scale, Security, Storage, Transparency, Trust, Ubiquity, Usability.

**Some attributes examples retrieved from IoT projects are:** Cost, Power, Size, Weight.

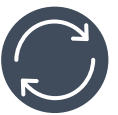

| | TO DO | DONE | N/A |
|---|---|---|---|
| Define what are the relevant attributes and their definition for the IoT project (such as "the voice command should always be available" - related to Availability, The ability of the service to be always available, regardless of hardware, software, or user fault). | ◉ | ◉ | ◉ |
| Define what are the measures and metrics for the selected attributes (Measure of Availability = uptime ÷ (uptime + downtime). | ◉ | ◉ | ◉ |
| Describe the implementation mechanisms for the selected attributes (such as using redundant infrastructure components for the voice command to ensure availability). | ◉ | ◉ | ◉ |
| Describe the observation and testing mechanisms for the selected attributes (such as the voice command will be tested monthly and should have 99% available time). | ◉ | ◉ | ◉ |

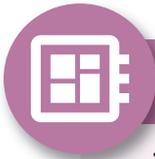

# THINGS

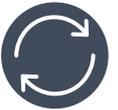

**1. Define and Implement components.** As a starting point in the IoT concept definition, considering the goals established from the problem domain is possible to extract the components required to achieve such a goal. After the components are identified, they all need to be defined with more detailed descriptions. After the components are identified and define, the components need to be implemented in the system. It is an ongoing activity, as the items should be revisited considering new information that can update the implementation, such as the environmental influence on a given component.

| | TO DO | DONE | N/A |
|---|---|---|---|
| Define components attributes (such as power, size, and memory). | ● | ● | ● |
| Describe the component's behavior (such as actuation, identification, monitoring, and sensing). | ● | ● | ● |
| Identify external partners (not internal to the system but are required for the solution). | ● | ● | ● |
| Identify components for interaction (such as traditional dashboard solutions or smartwatches and touch devices). | ● | ● | ● |
| Establish criteria for component selection (such as costs and restrictions). | ● | ● | ● |
| Describe a strategy for implementing and implementing necessary components (such as using microcontrollers like Arduino and Raspberry Pi, since they can provide a user-friendly development environment). | ● | ● | ● |
| Describe a strategy for adapting and adapt necessary components (such as wearables and aid for older adults that should be adjusted to the end-user). | ● | ● | ● |
| Describe a strategy for user customization (such as do-it-yourself philosophy using low-cost hardware and 3D printed parts). | ● | ● | ● |

**2. Define strategy for integration.** It can be a software layer (like middleware) or a physical layer (like circuit adapters) or another alternative that lies between the system components on each side as a bridge.

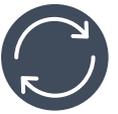

| | TO DO | DONE | N/A |
|---|---|---|---|
| Describe and implement a strategy to address integration issues (such as using interoperable open standards). | ◉ | ◉ | ◉ |
| Describe a strategy for integrating necessary components (such as using modular or layered architecture). | ◉ | ◉ | ◉ |
| Identify heterogeneity and incompatibility issues among components (such as check the communication technologies and protocols in use). | ◉ | ◉ | ◉ |

**3. Define device protection.** Related to the physical integrity of the components (system safety), like calibrate power. It is a responsibility to keep the system a state safe and not in danger or at risk (Cambridge Dictionary).

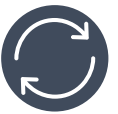

| | TO DO | DONE | N/A |
|---|---|---|---|
| Define physical threats among components (such as fires or flooding in the location of the components). | ◉ | ◉ | ◉ |
| Define the component's holder and integration needs (such as a combination of modules, a socket, or a device). | ◉ | ◉ | ◉ |
| Describe mechanisms to mitigate the threats (such as security measures to prevent physical damages). | ◉ | ◉ | ◉ |
| Describe and implement a strategy to address threats (such as Parental Control where changes can be made only by authorized individuals). | ◉ | ◉ | ◉ |

**4. Implement components identity.** From our IoT definition, the object should be uniquely identified, addressable. It should provide all the device identity information.

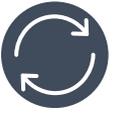

| | TO DO | DONE | N/A |
|---|---|---|---|
| Define management procedures (such as how to add or remove, enable or disable components from the system). | ● | ● | ● |
| Describe and implement device authentication (such as access control, to ensure that the system verifies the credentials). | ● | ● | ● |
| Describe and implement device identity (such as by IP address, with attributes and metadata defining physical or virtual identity). | ● | ● | ● |

**5. Define components temporality.** Uncertainties and issues related to temporality across the components should be addressed to reduce risks since they can be heterogenic.

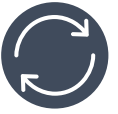

| | TO DO | DONE | N/A |
|---|---|---|---|
| Describe a strategy for real-time operation (such as real-time decision making and real-time monitoring). | ● | ● | ● |
| Describe a strategy for unifying system time across different components (such as unique timestamps). | ● | ● | ● |
| Describe a strategy for time-related quality attributes (such as availability and frequency). | ● | ● | ● |
| Indicate when the component performs its tasks (such as detail the sequence of activities related to a behavior). | ● | ● | ● |

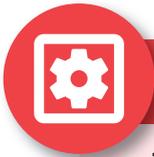

# BEHAVIOR

**1. Define identification.** The behavior of identifying things by labeling and enabling them to have an identity, recover (through reading), and broadcast information related to the thing and its state. It refers to physical identification - when objects are tagged with electronic tags containing specific information, making it possible to identify objects through tag readers. Not to be virtually identifiable in the sense of connectivity (e.g., IP address).

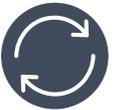

| | TO DO | DONE | N/A |
|---|---|---|---|
| Define the object to be identified (can be a car, a product or a person for example) | 🔴 | 🔴 | 🔴 |
| Define the metadata related to the object (id, name, and description for example) | 🔴 | 🔴 | 🔴 |
| Define an identification technology (QR code or RFID, for example) | 🔴 | 🔴 | 🔴 |
| Describe the reading event (Manual or automatic, for example) | 🔴 | 🔴 | 🔴 |
| Describe the type of identification (Static/Movel, Active/Passive, Disposable...) | 🔴 | 🔴 | 🔴 |

_______________________________________________

_______________________________________________

**2. Define sensing.** The primary function is to sense environment information, requiring information aggregation, data processing and transmission, controlling external context. Enables awareness, thus acting as a bridge between the physical and digital world.

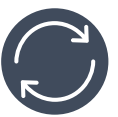

| | TO DO | DONE | N/A |
|---|---|---|---|
| Define the data related to the sensing (data to be extracted by the sensors with syntactic and semantic meaning...) | 🔴 | 🔴 | 🔴 |
| Describe a response for abnormal conditions (send an alert, activate actuation...) | 🔴 | 🔴 | 🔴 |
| Indicate the desired threshold and values (normal condition, safe values....) | 🔴 | 🔴 | 🔴 |
| Identify the sensing device (pressure , temperature sensor, Motion sensor...) | 🔴 | 🔴 | 🔴 |
| Establish the sensing rules (schedule-based sensing, event-based sensing, always-on sensing...) | 🔴 | 🔴 | 🔴 |

_______________________________________________

_______________________________________________

**3. Define actuation.** Mechanical interventions in the real world according to decisions based on aggregated data or even upon actors' right trigger; relay on responses to the collected information to perform actions in the physical world and change the object state.

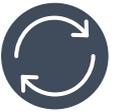

| | TO DO | DONE | N/A |
|---|---|---|---|
| Describe the manual or automatic mode (the use of rules, threshold, or response time can be applicable. In a smart farm, the irrigation is automatic according to the temperature.) | ● | ● | ● |
| Locate the action (In a smart farm, release water in the farm) | ● | ● | ● |
| Identify who triggers the action (device or human user, for example. In a smart farm, the farmer triggers the action.) | ● | ● | ● |
| Indicate the circumstances for triggering action - input (if the sensed date is below what is expected according to a defined threshold. In a smart farm, the irrigation is automatic when it is above 30 degrees C) | ● | ● | ● |
| Establish the consequences of an action - output (In a smart farm, the farm is irrigated) | ● | ● | ● |
| Identify who performs the action (device or human user, for example. In a smart farm, the smart houses connected to the water tank) | ● | ● | ● |

**4. Define monitoring.** A solution to watch, keep track of, or constantly check for a special purpose (observing without control).

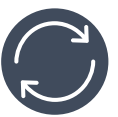

| | TO DO | DONE | N/A |
|---|---|---|---|
| Define data to be monitored (environmental or healthy information for example) | ● | ● | ● |
| Describe monitoring rules (such as where to send information, alerts, and flags) | ● | ● | ● |
| Identify the monitoring device (sensor, tag...) | ● | ● | ● |
| Indicate the monitoring temporality (real-time, once a day, during summer...) | ● | ● | ● |

**5. Define user behavior.** Elements that interacts with the user, including the devices that enable data visualization or voice commands.

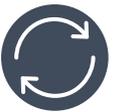

|  | TO DO | DONE | N/A |
|---|---|---|---|
| Describe and define data to be shared (to make human sense from the data received, such as by using dashboards or sonore alarms) | ⦿ | ⦿ | ⦿ |
| Define and establish interaction rules (from what is received, what the user can do next) | ⦿ | ⦿ | ⦿ |
| Define interaction devices and Identify their roles (mobile app, gesture recognition or smartwatch...). | ⦿ | ⦿ | ⦿ |

_______________________________________________
_______________________________________________

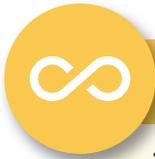

# INTERACTIVITY

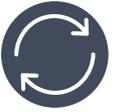

**1. Define involved actors.** Identify any human, object or thing that engages in an interaction with the system, including other systems.

| | TO DO | DONE | N/A |
|---|---|---|---|
| Define system admin and responsibilities. (Such as, who is responsible for updates). | ⬤ | ⬤ | ⬤ |
| Define the users, roles and responsibilities (Consider user, business, legal, regulatory and functional issues: for example, requirements for special needs). | ⬤ | ⬤ | ⬤ |
| Describe and Establish user control of configurations, rules and generated data. (Such as, settings of timers and alarms or authorization for shared data). | ⬤ | ⬤ | ⬤ |
| Define safety procedures for human users. (Such as access to the physical device by biometric control). | ⬤ | ⬤ | ⬤ |
| Describe and Establish the data personalization per user/role (For example, access control solutions for both the users and components where certain actions can only be associated with a specific role). | ⬤ | ⬤ | ⬤ |

**2. Define Interaction Methods.** IoT innovate the interactions perspectives the things can engage in Human-Thing (HTI) and Thing-Thing interaction (TTI). HTI is related to human users, and the things, any object that the user will interact with and that has enhanced behaviors through software. TTI refers to the interactivity and interoperability between the things themselves, in varying forms.

Interaction object (related to things): Input devices: including any type of component acting as bridge for interaction between actor and the system. **Output devices:** referring to the environment ''devices'' that act as actuators and provide results and information.

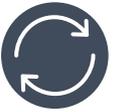

| | TO DO | DONE | N/A |
|---|---|---|---|
| Define and implement interaction method (Such as gesture and gaze, voice and audio, touch and tactile, traditional GUI, or multi-method with a combination of these) | ◉ | ◉ | ◉ |
| Identify interaction object (For gestures for example, the movements are acquired from camera streams by using computer vision techniques) | ◉ | ◉ | ◉ |
| Define and Establish interaction grammar (For gestures for example, the grammar is a set of know gestures and movements supported by the system like Up, Down, Left, Right, Forward, Backward) | ◉ | ◉ | ◉ |
| Define and Establish interaction recognition (For gestures, is the component to identify and process what gesture or movement the user is doing by using Dynamic Time Warping algorithm for example) | ◉ | ◉ | ◉ |
| Identify interaction sequence and expected result (such as the action sequence between user and system to gather sensor information) | ◉ | ◉ | ◉ |

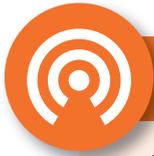

# CONNECTIVITY

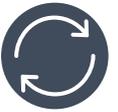

**1. Establish Connectivity.** For dynamic linking IoT services is necessary compatible connectivity based on topology, architecture, constraints and standards.

|  | TO DO | DONE | N/A |
|---|---|---|---|
| Define network topology and architecture (such as how the connection is organized, node-to-node communication through two active devices with NFC). | ◉ | ◉ | ◉ |
| Define connectivity constraints (Considering the systems requirements, define connectivity constraints such as frequency, range, nodes, power, data rate). | ◉ | ◉ | ◉ |
| Define and Implement connectivity standards (to enable the system to operate sharing the same environment by using LoWPAN , BTv5 or ZigBee...). | ◉ | ◉ | ◉ |
| Establish Service Discovery mechanisms. Different IoT solutions can require different properties to identify suitable services to mash-up (for example, use semantic-based similarity and quality of service). | ◉ | ◉ | ◉ |

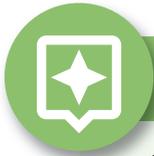

# SMARTNESS

**1. Define and Implement smartness.** Smartness deals with the combination of characteristics that enable the IoT system to be semi or entirely autonomous for performing any action in the environment. The actions are associated with the smartness ability, depending on the application domain and the user's needs. These characteristics of smartness are systems requirements. The data collected from the environment supports the system to be aware, decide, and act. Therefore smartness should be defined according to the user's need, combining Environment, Data, and Behavior facets.

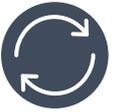

|  | TO DO | DONE | N/A |
|---|---|---|---|
| Describe and Implement AI technology (ex. Machine learning, Fuzzy logic). | ● | ● | ● |
| Describe and implement data processing (ex. Required analysis for interpretation or management of component, data, and interaction). | ● | ● | ● |
| Describe and implement data semantics (ex. data interpretation and ontologies). | ● | ● | ● |
| Describe a strategy for real-time operation (Real-time decision, real-time monitoring, or real-time visualization). | ● | ● | ● |
| Identify the decision-makers (ex. users, software system). | ● | ● | ● |

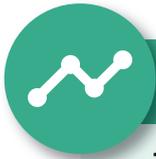

# DATA

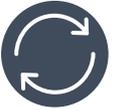

**1. Define and Implement the data model.** The project should provide the modeling of the data sources in the system definition. It should capture the relationships existing between a source and the physical environment and the relationships existing among data sources themselves. It is also used to specify the properties and characteristics of the retrieved data. Data has great value for the IoT systems, and it is as relevant as the sources defined for the project in question. In a world of possibilities, it is necessary precision and adequacy to determine relevant data.

| | TO DO | DONE | N/A |
|---|:---:|:---:|:---:|
| Define the properties (such as metadata, data types required to achieve the tasks at hand) | ◎ | ◎ | ◎ |
| Define what data will be collected from the components, users and external systems (volume, variety, speed, value, static, dynamic) | ◎ | ◎ | ◎ |
| Define the inputs and outputs applicable for each data source (assure that the devices or systems are getting and generating accurate data) | ◎ | ◎ | ◎ |
| Describe the relationships and flows (such as express relations among the components) | ◎ | ◎ | ◎ |
| Describe how data will be used (related to the system behavior and the goals defined in the problem domain) | ◎ | ◎ | ◎ |
| Describe how data will be collected from each source (such as cloud, Bluetooth...) | ◎ | ◎ | ◎ |
| Describe how data will be shared (such as cloud, Bluetooth...) | ◎ | ◎ | ◎ |
| Describe and implement the data model (Conceptual, Physical and Logical data modeling) | ◎ | ◎ | ◎ |
| Identify the data sources (such as sensors, actuators, external partners...) | ◎ | ◎ | ◎ |

**2. Implement data protection and privacy.** While defining the data required, it is necessary to provide users' consent, privacy preferences, and comply with regulations (GDPR, Data Protection Act 2018....)

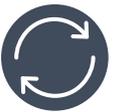

| | TO DO | DONE | N/A |
|---|---|---|---|
| Describe a strategy to ensure privacy (For example, define what is considered 'private data' that can impact a human's privacy including the combination of data and anonymize any personal information extracted or inferred, or granting each endpoint unique tokens). | ◉ | ◉ | ◉ |
| Describe a strategy for user control over their data (such as obtain the user's consent, where required and easily express their privacy preferences in a 'permissions dashboard'). | ◉ | ◉ | ◉ |
| Describe a strategy for risks and weakness related to data (For example, keep libraries updated and monitors the databases). | ◉ | ◉ | ◉ |
| Describe and implement data classification (For example, Content-based classification based on the interpretation of files and sensitive information). | ◉ | ◉ | ◉ |
| Establish and address law requirements and regulations (Data protection laws such as the EU GDPR). | ◉ | ◉ | ◉ |
| Establish the risks and vulnerability related to data (For example, Information modification, Denial-of-service attacks, and service interruption). | ◉ | ◉ | ◉ |

**3. Define data temporality.** The environment can change over time. For this reason it is important to have accurate, update and valid data. Each data source can be independently integrated and heterogenic, therefore issues related to data temporality across the sources should be addressed to reduce risks.

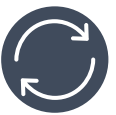

| | TO DO | DONE | N/A |
|---|---|---|---|
| Define a capture frequency (For example, the data collection process can be done once a day). | ◉ | ◉ | ◉ |
| Define a expiration procedure (For example, the capability to auto-expire after a full capture cycle). | ◉ | ◉ | ◉ |
| Define a strategy for data removal (For instance, at the end of the data lifecycle). | ◉ | ◉ | ◉ |

**4. Provide data storage.** Data storage is the recording (storing) of information (data) in a storage medium. It is related to cloud and edge solution, in-memory caches, temporary or permanent archives.

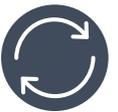

| | TO DO | DONE | N/A |
|---|---|---|---|
| Describe a security strategy for the data stored (For example, key management, and authentication mechanisms) | ⦿ | ⦿ | ⦿ |
| Describe a strategy for data related quality attributes (availability, performance, scalability and other) | ⦿ | ⦿ | ⦿ |
| Describe a strategy for data storage (Such as cloud, edge, local servers...) | ⦿ | ⦿ | ⦿ |

**5. Implement aggregation, synchronization and conflict resolution.**

**Aggregation:** For example, gather and summarize data from multiple devices. This may be relevant when having several connected IoT devices with a large amounts generated data that could lead to heterogeneous traffic loads, data redundancy and energy consumption.

**Synchronization:** For example, when different devices are added or removed from an environment it can generate inconsistent data and lead devices to lose sync with each other.

**Conflict Resolution:** For example, different requests can occur with opposite goals, like when an user may want to activate the climate control before arriving at home via smartphone, but the system deactivated it because no one is at home.

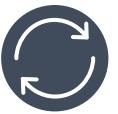

| | TO DO | DONE | N/A |
|---|---|---|---|
| Describe a strategy for data aggregation (A possible solution can be Cluster-based or chain-based aggregation methods) | ⦿ | ⦿ | ⦿ |
| Describe a strategy for data synchronization and Indicate when it is applied (A possible solution is to implement Multiple-round synchronization techniques, where the systems is synchronized by blocks of data, sended in rounds). | ⦿ | ⦿ | ⦿ |
| Describe a strategy for conflict resolution and Indicate when it is applied (A possible solution is to implement rules so the human users have the authority to cancel, postpone or redo actions in the event of a conflict.). | ⦿ | ⦿ | ⦿ |

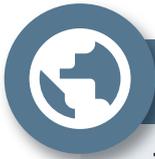

# ENVIRONMENT

**1. Define relevant environment information.** The environment where the solution is deployed is a multi-dimensional contextual space with different levels of importance that can change over time. It is necessary to state the contextual variables to be used to translate the environment into computing technologies when considering the context. Systems can adapt their behavior according to the information they receive about the environment or the user's, and this information is the context the systems should be aware of.

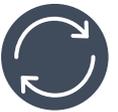

|  | TO DO | DONE | N/A |
|---|:---:|:---:|:---:|
| Define what data will be collected from the environment (for example, temperature, and humidity). | ◉ | ◉ | ◉ |
| Describe how to collect data from the environment (for example, using RFID and sensors). | ◉ | ◉ | ◉ |
| Describe the context of use. The Context of use includes: i) user, with all needs as well as specific abilities and preferences; ii) environment, in which interaction occurs; and iii) IoTsystem, composing of hardware and software | ◉ | ◉ | ◉ |

**2. Define the environmental impact.** The interplay between environment and solution can affect on each other that can alter the desired outcome, which we should be aware of.

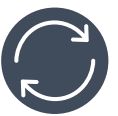

|  | TO DO | DONE | N/A |
|---|:---:|:---:|:---:|
| Describe a strategy for the issues observed (ex. physical protection) | ◉ | ◉ | ◉ |
| Establish the environmental impact in the solution (physical - ex. deployed in the water) | ◉ | ◉ | ◉ |
| Establish the environmental impact in the solution through time (physical - ex. when it rains) | ◉ | ◉ | ◉ |
| Establish the environmental impact on the quality of the solution (virtual - ex. city noise crashing with the building quality) | ◉ | ◉ | ◉ |
| Establish the solution impact in the environment (physical - ex. birds living nearby) | ◉ | ◉ | ◉ |

**3. Control the physical access to the solution.** Just like virtual protection (security), it is necessary to control physical access to the solution for some cases.

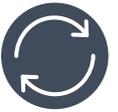

| | TO DO | DONE | N/A |
|---|---|---|---|
| Describe the mechanisms for control and Identify affected roles (ex.: passwords). | ◉ | ◉ | ◉ |
| Describe the mechanisms for unauthorized access (ex.: alarm) | ◉ | ◉ | ◉ |

**4. Define Digital Environment.** Solutions involving Augmented Reality, Immersion and Simulation, Holograms and 3D Digital Interaction should be defined and implemented.

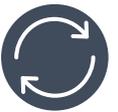

| | TO DO | DONE | N/A |
|---|---|---|---|
| Define and Establish the digital environment (ex.: speaking holograms activated by motion sensors in a museum) | ◉ | ◉ | ◉ |

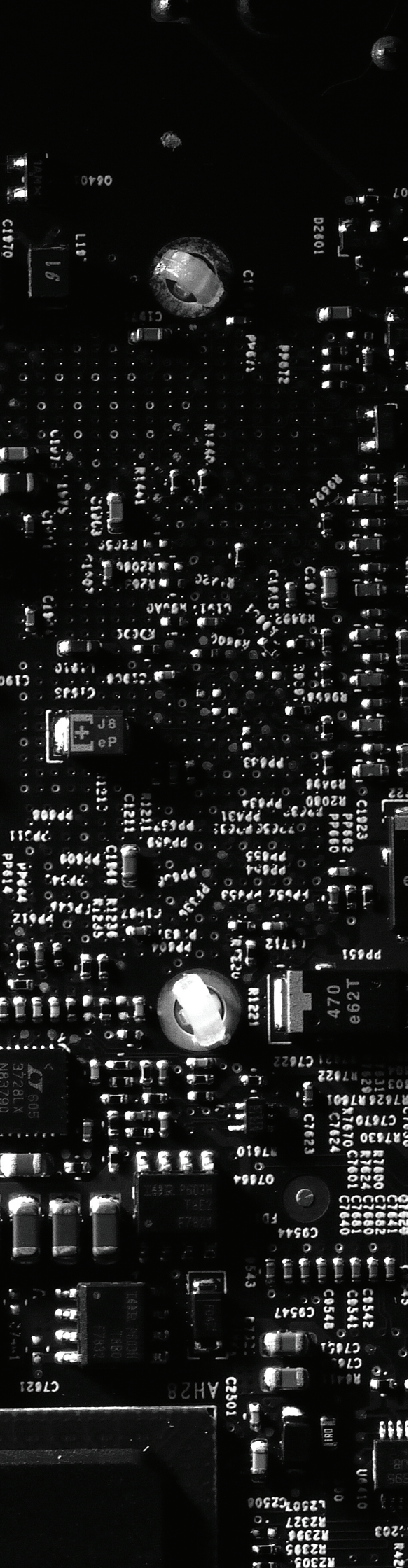

# SUMMARY

By the end of this IoT Roadmap, the project team should be able to have a clear understanding of the problem domain, the operational requirements within, and the conditions and constraints for the solution.

# RESULTS

The marks on the Roadmap indicates:

| ITEMS TO DO | ITEMS DONE | ITEMS NOT APPLICABLE |
|---|---|---|
|  |  |  |

# NEXT STEPS

Depending on the project plan, take the opportunity to revisit Cross Cutting items and make the necessary adjustments in the engineering process.

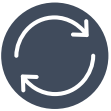

If the product is already completed, consider handling specific items for the IoT solution, such as:

• Documentation, installation and update procedures (related to maintainability, should include maintenance and replacement orientations).

• Procedures for remote boot and physical boot (for example, how to make a new device registration and the first connection).

• Recovery and contingency plan (it can start by establishing vulnerability action procedures).

# CONCLUSION

We hope the Roadmap can contribute to the understanding of IoT and its related information. By using the IoT Roamap to support project decisions to perceive and handle needs, demands and risks associated with engineering an IoT solution.

For questions or suggestions, please contact: rmotta@cos.ufrj.br

Related institutions:

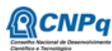 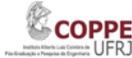 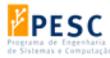 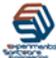 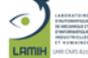 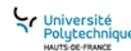 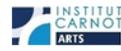

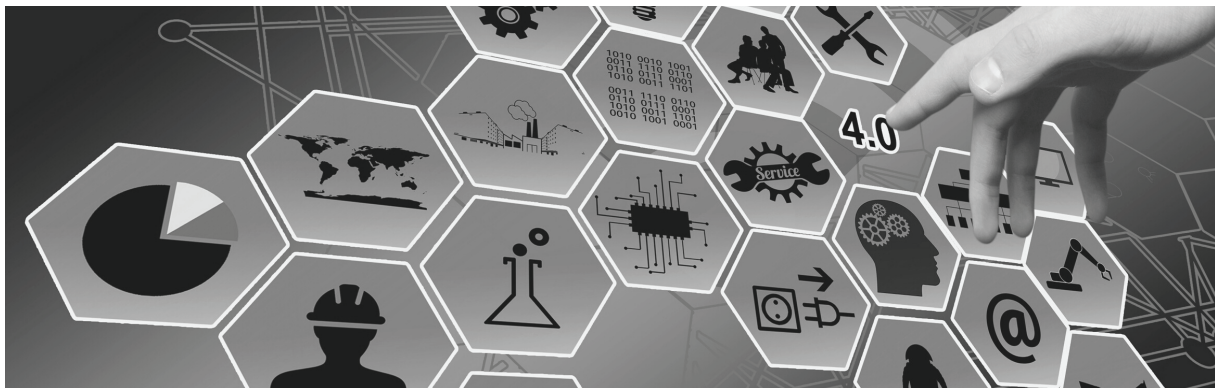

# FURTHER INFORMATION

**About IoT:**

• Motta, R. C.; da Silva, V. M.; Travassos, G. H. "Towards understanding of IoT". Link: http://doi.org/10.5753/jserd.2019.14

• Motta, R. C., de Oliveira, K. M. , Travassos, G. H. "On challenges in engineering IoT systems." Link: http://doi.org/10.5753/jserd.2019.15

• Motta, R. C., de Oliveira, K. M. ,Travassos, G. H. "A Framework to Support the Engineering of Internet of Things Software Systems." Link: http://doi.org/10.1145/3319499.3328239

**About Quality Characteristics:**

• ISO/IEC 25000:2014 Systems and software engineering - Systems and software Quality Requirements and Evaluation (SQuaRE) Link: https://www.iso.org/standard/64764.html

**About Supporting Technologies:**

• de Souza, B. P., Motta, R. C., Travassos, G. H. "Towards the Representation of Smartness in IoT Scenarios Specification" Link: http://doi.org/10.1145/3350768.3351797

• de Souza, B. P., Motta, R. C., Travassos, G. H. "The first version of SCENARIotCHECK: A Checklist for IoT based Scenarios." Link:http://doi.org/10.1145/3350768.3350796

• da Silva, V. "SCENARIoT Support for Scenario Specification of Inte net of Things-Based Software Systems" Link: http://zenodo.org/record/4080680

**About Systems Engineering:**

• SEBoK contributors, "Guide to the Systems Engineering Body of Knowledge", May 2020. Link: https://www.sebokwiki.org/